\documentclass[letterpaper,twocolumn,prl,aps,superscriptaddress,amsmath,amssymb,floatfix]{revtex4-1}
\usepackage{mathptmx}
\usepackage[latin9]{inputenc}
\usepackage{color}
\usepackage{verbatim}
\usepackage{float}
\usepackage{amsmath}
\usepackage{amssymb}
\usepackage{graphicx}
\usepackage{esint}
\usepackage{siunitx}

\usepackage[unicode=true,
 bookmarks=true,bookmarksnumbered=false,bookmarksopen=false,
 breaklinks=false,pdfborder={0 0 1},backref=false,colorlinks=true]
 {hyperref}
\hypersetup{
 linkcolor=magenta,urlcolor=blue,citecolor=blue,pdfstartview={FitH},hyperfootnotes=false}

\makeatletter

\usepackage{textcomp}
\usepackage{epstopdf}

\pdfpageheight\paperheight
\pdfpagewidth\paperwidth

\@ifundefined{textcolor}{}{%
 \definecolor{BLACK}{gray}{0}
 \definecolor{WHITE}{gray}{1}
 \definecolor{RED}{rgb}{1,0,0}
 \definecolor{GREEN}{rgb}{0,1,0}
 \definecolor{BLUE}{rgb}{0,0,1}
 \definecolor{CYAN}{cmyk}{1,0,0,0}
 \definecolor{MAGENTA}{cmyk}{0,1,0,0}
 \definecolor{YELLOW}{cmyk}{0,0,1,0}
}

\usepackage{xcolor}
\usepackage{soul}
\definecolor{blue}{rgb}{0,0,1}
\definecolor{red}{rgb}{1,0,0}
\definecolor{green}{rgb}{0,1,0}

\usepackage{soul}

\makeatother

\begin{document}

\title{Controllable atomic collision in a tight optical dipole trap}

\author{Zhu-Bo~Wang}
\thanks{These two authors contributed equally to this work.}
\affiliation{CAS Key Laboratory of Quantum Information, University of Science and Technology of China, Hefei 230026, China}
\affiliation{CAS Center For Excellence in Quantum Information and Quantum Physics, University of Science and Technology of China, Hefei 230026, China}

\author{Chen-yue Gu}
\thanks{These two authors contributed equally to this work.}
\affiliation{CAS Key Laboratory of Quantum Information, University of Science and Technology of China, Hefei 230026, China}
\affiliation{CAS Center For Excellence in Quantum Information and Quantum Physics, University of Science and Technology of China, Hefei 230026, China}

\author{Xin-Xin Hu}
\affiliation{CAS Key Laboratory of Quantum Information, University of Science and Technology of China, Hefei 230026, China}
\affiliation{CAS Center For Excellence in Quantum Information and Quantum Physics, University of Science and Technology of China, Hefei 230026, China}
\author{Ya-Ting Zhang}
\affiliation{CAS Key Laboratory of Quantum Information, University of Science and Technology of China, Hefei 230026, China}
\affiliation{CAS Center For Excellence in Quantum Information and Quantum Physics, University of Science and Technology of China, Hefei 230026, China}
\author{Ji-Zhe Zhang}
\affiliation{CAS Key Laboratory of Quantum Information, University of Science and Technology of China, Hefei 230026, China}
\affiliation{CAS Center For Excellence in Quantum Information and Quantum Physics, University of Science and Technology of China, Hefei 230026, China}
\author{Gang Li}
\affiliation{State Key Laboratory of Quantum Optics and Quantum Optics Devices, and Institute of Opto-Electronics, Shanxi University, Taiyuan 030006, China}
\affiliation{Collaborative Innovation Center of Extreme Optics, Shanxi University, Taiyuan 030006, China.}
\author{Xiao-Dong He}
\affiliation{State Key Laboratory of Magnetic Resonance and Atomic and Molecular Physics, Innovation Academy for Precision Measurement Science and Technology, Chinese Academy of Sciences, Wuhan 430071, China}
\author{Xu-Bo Zou}
\affiliation{CAS Key Laboratory of Quantum Information, University of Science and Technology of China, Hefei 230026, China}
\affiliation{CAS Center For Excellence in Quantum Information and Quantum Physics, University of Science and Technology of China, Hefei 230026, China}
\author{Chun-Hua Dong}
\affiliation{CAS Key Laboratory of Quantum Information, University of Science and Technology of China, Hefei 230026, China}
\affiliation{CAS Center For Excellence in Quantum Information and Quantum Physics, University of Science and Technology of China, Hefei 230026, China}
\author{Guang-Can Guo}
\affiliation{CAS Key Laboratory of Quantum Information, University of Science and Technology of China, Hefei 230026, China}
\affiliation{CAS Center For Excellence in Quantum Information and Quantum Physics, University of Science and Technology of China, Hefei 230026, China}
\author{Chang-Ling Zou}
\email{clzou321@ustc.edu.cn}
\affiliation{CAS Key Laboratory of Quantum Information, University of Science and Technology of China, Hefei 230026, China}
\affiliation{CAS Center For Excellence in Quantum Information and Quantum Physics, University of Science and Technology of China, Hefei 230026, China}
\affiliation{State Key Laboratory of Quantum Optics and Quantum Optics Devices, and Institute of Opto-Electronics, Shanxi University, Taiyuan 030006, China}

\date{\today}
\begin{abstract}
Single atoms are interesting candidates for studying quantum optics and quantum information processing. Recently, trapping and manipulation of single atoms using tight optical dipole traps have generated considerable interest. Here we report an experimental investigation of the dynamics of atoms in a modified optical dipole trap with a backward propagating dipole trap beam, where a change in the two-atom collision rate by six times has been achieved. The theoretical model presented gives a prediction of high probabilities of few-atom loading rates under proper experimental conditions. This work provides an alternative approach to the control of the few-atom dynamics in a dipole trap and the study of the collective quantum optical effects of a few atoms.
\end{abstract}

\maketitle

In the last few decades, attractive research directions have been unlocked by experimental achievements in single-atom trapping~\cite{Schlosser2001,Schlosser2002,Xu2010,Fung2015,Grunzweig2010a,Kuppens2000}, cooling~\cite{Thompson2013,Chin2017}, and manipulation, such as rearrangement~\cite{Endres2016,Barredo2016,Sheng2021} and transport~\cite{Kuhr2001,Dorevic2021}. Benefiting from the high-precision manipulation of atomic internal and external degrees of freedom, the use of single atoms provides a unique tool to explore the fundamental atom-atom interaction~\cite{Rui2017} and molecular formation~\cite{He2020}. In far-off-resonant dipole traps, also known as atom-tweezers, single atoms are arranged, which can serve as individual quantum bits in potential applications in  quantum simulation~\cite{Semeghini2021}, quantum computation~\cite{Graham2022,Li2019}, and quantum metrology~\cite{Norcia2019}.

The atomic collisional blockade effect~\cite{Schlosser2002,Fung2015} occurs in a dipole trap when the trapping volume is extremely small, which prevents the loading of multiple atoms, leading to a high loading rate of single atoms. The control of the collision process, making the high-probability loading of a few atoms possible, is also of interest. Two methods have been proposed before. By changing the size of the dipole trap~\cite{Schlosser2002,Kuppens2000} or using blue-detuned light to assist the elastic collision process~\cite{Sortais2012,Grunzweig2010a}, control of the collision loss rate can be realized. This is important in the realization of few-atom systems, which are highly desired for the fundamental study of collective light-matter interactions~\cite{Glicenstein2020,Ebert2014} and potential applications in quantum information coding~\cite{Pedersen2009a}.

\begin{figure}[h]
\centering
\includegraphics[width=\linewidth]{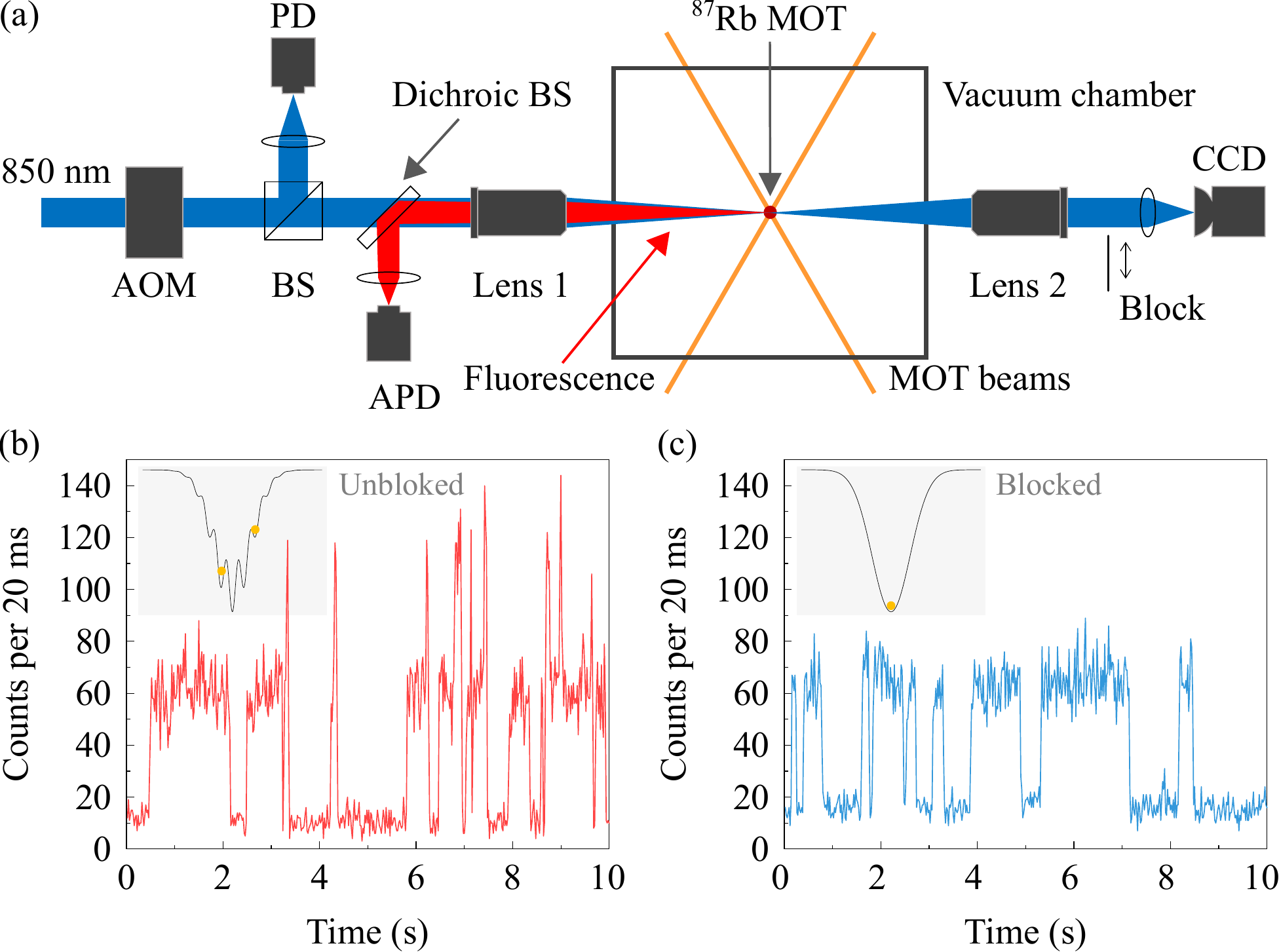}
\caption{(a) The experimental setup of a tight optical dipole trap for single atoms. PD: photodiode, AOM: acousto-optic modulator, BS: beam splitter, APD: avalanche photodiode, MOT: magneto-optical trap. Typical traces of the photon counting rate collected by the APD when the block between Lens~2 and CCD is removed and present are shown in (b) and (c), respectively. The insets illustrate the modification of the dipole trap due to the reflected dipole laser along the axial direction.%(c)The second-order correlation function of the atomic fluorescence when the trap beam is blocked.
}
\label{fig1}
\end{figure}

In this work, we demonstrate a new method to control the collision between atoms in a dipole trap experimentally, which extends single-atom trapping to multiple-atom trapping. By introducing a counter-propagating dipole laser beam, we modulate the trap, creating multiple local potential minima and hence reducing the chance of two atoms meeting together. With this approach, the average coexistence time of two $^{87}$Rb atoms in a single trap increases from $\SI{5.5}{\ms}$ to $\SI{29.9}{\ms}$. The collision rate is quantitatively investigated with a Monte Carlo model by numerically fitting the lifetimes of zero-, single- and two-atom events. Our model also predicts that in a single dipole trap, by properly tuning the collision rate and the density of the atom cloud, the loading probability of three or four atoms can exceed $20\%$ with lifetimes longer than $\SI{10}{\ms}$. This work provides a new technique to investigate and control cold-atom dynamics, which offers new opportunities in studying the collective interaction between atoms and light.

% with a $1/e^{2}$ radius of $w_{0}=\SI{2.1}{\um}$ and a trap depth of \SI{1.3}{\milli\kelvin}
%mathematical model to characterize single atom's loading and loss process in a tight red-detuned dipole trap. Based on this model, we put forward a method to calculate three parameters in the model which are loading rate, single loss rate and collision loss rate from experimental atomic fluorescence data. We use a Monte Carlo simulation to test this method's accurateness. We find a new way to control the collision loss rate by reflecting the trap laser back to the original path.

The experimental setup is illustrated in Fig.~\ref{fig1}(a). The dipole trap is formed with a far-off-resonant red-detuned laser (\SI{850}{nm}, \SI{36.0}{\mW}), which is focused to a $1/e^{2}$ radius of $w_{0}=\SI{2.2}{\um}$ by an objective lens [Lens~1 in Fig.~\ref{fig1}(a), $\mathrm{N.A.}=0.23$], causing an AC Stark shift of \SI{23.0}{\MHz} to the $F=2$ ground state~\cite{Shih2013}. The power of the dipole laser is stabilized to $\pm0.2\%$ (in $2$ hours) via an AOM that uses the PD output as the error signal for feedback. The dipole trap, with radial and axial trap frequencies of \SI{47.0}{\kHz} and \SI{4.1}{\kHz}, respectively, overlaps with a cloud of $^{87}$Rb atoms created by the MOT.
%, which is composed of six counter-propagating $\sigma^{+}$-$\sigma^{-}$ polarized cooling lasers, each containing a small amount of repumping laser, and a gradient magnetic field of \SI{10}{G/\cm} produced by a pair of anti-Helmholtz coils. Each cooling laser beam has an average intensity of \SI{1.9}{\mW\per\square\cm} and is \SI{12}{\MHz} red detuned to the $F=2$ to $F'=3$ $D2$ transition, while the repumping laser is resonant to the $F=1$ to $F'=2$ $D2$ transition.
The fluorescence of the atom(s), stimulated by the cooling lasers that are near-resonant to the D2 transitions, will be collected by Lens~1 and then detected by the APD. %after passing through a band-pass filter and a single-mode fiber(not shown in the figure) to eliminate stray noise.
An additional Lens~2 is used to ensure that the \SI{780}{\nano\metre} collection position overlaps with the center of the dipole trap.

According to the atomic collisional blockade effect~\cite{Schlosser2002,Fung2015}, when two atoms are captured in a tight optical dipole trap, both of them will collide out due to the red-detuned laser probe, thus the dipole trap can only load a single atom or zero atom. However, the signal that represents two-atom loading are observed in our setup rather frequently. A typical trace of atomic fluorescence of the dipole trap is presented in Fig.~\ref{fig1}(b), where three stages of photon counting rate are shown in the experiments. This indicates that two or even more atoms can exist for an observable time in the dipole trap. Further experimental investigation reveals that the two-atom events are obtained only when the backward propagating dipole trap beam exists. Once we block the optical path of the CCD, the third stage of the photon counting rate disappears, as shown in Fig.~\ref{fig1}(c).

In the experiments, the trap beam is adjusted to be nearly perpendicularly transmitted to the CCD sensor whose reflectivity (Sony IMX174 in Basler acA1920-155um) is measured as $24\%$. We conjecture that the interference between the forward and backward propagating trap beams modifies the dipole trap to be as a standing-wave-like trap [insets in Fig.~\ref{fig1}(b)]. The local potential minima obstruct the traveling of atoms along the dipole laser beam direction in the trap (with all other conditions preserved), leading to a decrease in the collision rate between atoms. A higher loading rate of multiple atoms in the dipole trap is hence achieved. It is anticipated that if the modulation of the potential is strong enough, the minima could be treated as individual traps for single atoms, as demonstrated in optical lattices~\cite{Bloch2008}. In contrast, when the modulation is weak, the dipole trap should be treated as a single trap, where atoms with certain temperature can still travel among the minima but with effective resistance along the axial direction. In our experimental setup, assuming that the alignment between the forward and backward dipole laser beams is perfect, a backward beam with a power of \SI{400}{\uW} can induce a modulation of potential of $\sim\pm20\%$. Thus, instead of being formed as a lattice of traps, the dipole trap is expected to be slightly modified under this condition.

\begin{figure}[h]
\centering
\includegraphics[width=\linewidth]{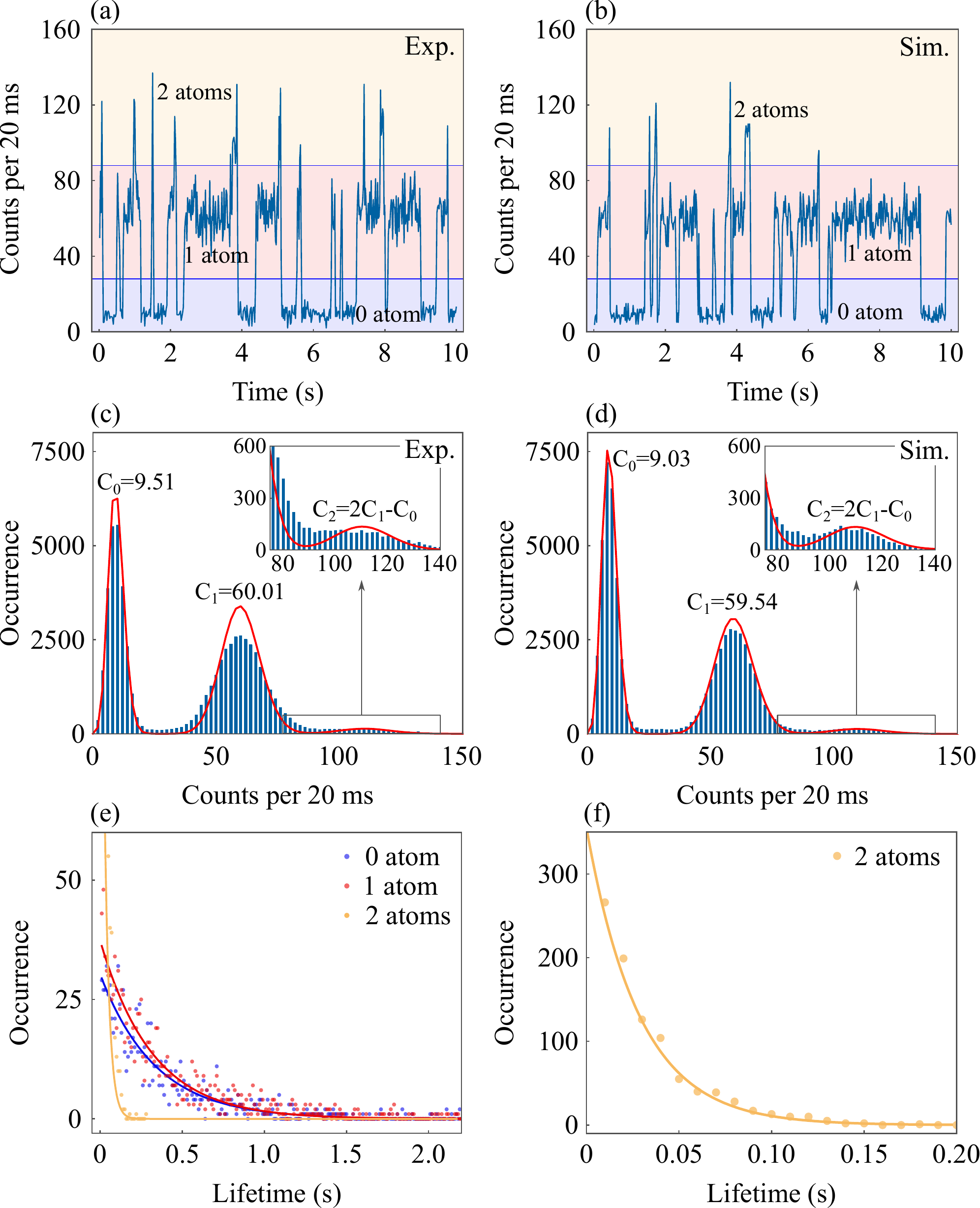}
\caption{(a) and (b) Typical traces of the photon counting rate due to the fluorescence of trapped atoms from the experiment and simulation, respectively. (c) and (d) Histograms of the counting rate accumulated in 20 minutes from the experiment and simulation, respectively. The red line is a fitting with three Poisson distributions, where $C_{n}$ denotes the average counting rate of the $n$-atom state. Insets are the enlarged plots for the 2-atom states. (e) Occurrence of the lifetime of the 0-, 1- and 2-atom states in the trap from the experimental results. A zoomed-in plot for the 2-atom state is shown in (f).
}
\label{fig2}
\end{figure}

To quantify the influence of the back-reflected dipole laser on the collision rate, we develop a theoretical model based on Ref.~\cite{Schlosser2002,Fung2015} to numerically simulate the atom loading and loss process, with which the corresponding rates can be extracted. Considering the number of atoms in the dipole trap as a state of a discrete-time Markov chain, and assuming that only one process occurs in one step with a short enough period of time ($\tau$), the probabilities of atomic number change in one step can be written as
\begin{equation}
 \begin{split}
 P_{n\to n+1}&=R\tau, \\
 P_{n\to n-1}&=n\gamma\tau, \\
 P_{n\to n-2}&=n(n-1)\beta\tau/2,\\
 P_{n\to n}&=1-P_{\tau,n\to n+1}-P_{\tau,n\to n-1}-P_{\tau,n\to n-2}.
 \end{split}
\label{transition}
\end{equation}
Here, the subscript $n\to n^{\prime}$ represents the transition from $n$ to $n^{\prime}$ atoms in the dipole trap. The coefficients of single-atom loading, single-atom loss and two-atom collision loss rate are denoted by $R$, $\gamma$ and $\beta$, respectively. A typical trace of the experimental observation of the atom loading and loss is shown in Fig.~\ref{fig2}(a). The photon counting rate of fluorescence increases tremendously when one atom is captured and decreases when the atom escapes. From the experimental traces, we plot the occurrence of the photon counts, which gives three Poisson distribution profiles, indicating the zero-atom ($C_{0}=9.51$), single-atom ($C_{1}=60.01$) and 2-atom ($C_{2}=110.51$) states in the dipole trap, as presented in Fig.~\ref{fig2}(c). Since the two-atom events are rare, we deduce the value of counts per \SI{20}{\milli\second} by $C_{2}=2C_1-C_0$ with the assumption that the photon emission rate is proportional to the atomic number. According to the fitting of the Poisson distribution of the photon counts, shown as the red curve in Fig.~\ref{fig2}(c), we set the thresholds [28 and 88, blue horizontal lines in Fig.~\ref{fig2}(a)] to discriminate zero-, single- and two-atom events. When the fluorescence counts are lower than 28 counts per \SI{20}{\milli\second}, zero atom is captured, while with counts between 28 and 88 (higher than 88), the trap is considered in the single- (two-) atom state.

To extract $R$, $\gamma$ and $\beta$ from the experiment, we calculate the probability of the dipole trap being in the zero-, single- and two-atom states as a function of the lifetime $t_{0}$ of atom(s) staying in the trap:
\begin{equation}
    \label{Eq_propability}
    P_{t_0,n}=(P_{\tau,n\to n})^{m}\overset{\lim\limits_{\tau\to 0}}{=}e^{-(R+n\gamma+\frac{n(n-1)}{2}\beta)t_0}, %\label{lifetime}
\end{equation}
where $t_0=m\tau$. Under an approximation that considers all the states with more than two atoms as the two-atom state, the decay rates of the lifetimes of the zero-, single- and two-atom states are $R$, $R+\gamma$ and $2\gamma+\beta$ respectively. We can then derive $R$, $\gamma$, and $\beta$ from experimental data with the decay rates of the lifetimes of the trap's three different states. By fitting the experimental decay rate of each state, we obtain $R=2.99\pm0.15~\mathrm{s}^{-1}$, $\gamma=0.16\pm0.20~\mathrm{s}^{-1}$ and $\beta=34.55\pm0.56~\mathrm{s}^{-1}$, as shown in Figs.~\ref{fig2}(e) and (f).% Note that a negative $\gamma$ is not attainable in physics, which is attributed to a small enough single-atom loss rate $\gamma$ and a restricted precision.

To verify the accuracy of this model, we carried out numerical simulations based on Eq.~\ref{transition}. In the simulation, to imitate practical experiments, for each step with a time of $\tau=\SI{1}{\us}$ in the Monte Carlo simulation, we generate a random number between zero and one to decide whether a photon is emitted by the atom and then detected by the APD in this step. Specifically, if the random number is smaller than $C_{n}\tau$, one photon is collected, otherwise no photon is collected. A typical simulation trace of the photon counting rate and its occurrence are shown in Figs.~\ref{fig2}(b) and (d), respectively, with all parameters adapted from experiments. From the simulation results, we obtain the corresponding coefficients $R_{s}=2.83\pm0.13~\mathrm{s}^{-1}$, $\gamma_{s}=0.08\pm0.17~\mathrm{s}^{-1}$ and $\beta_{s}=35.15\pm0.56~\mathrm{s}^{-1}$, which are in high agreement with the initial input parameters. This validates the effectiveness of our model in describing the dynamics of the atoms' behavior in this dipole trap. A difference between the experiment and the simulation is the step time $\tau$. In the simulation, $\tau$ is chosen to be $\SI{1}{\us}$ to ensure that only one process can occur in one step. In experiments, $\tau$ needs to be $\SI{20}{\ms}$ to obtain a high fidelity to distinguish between different atomic numbers. When $R\tau$ or $\gamma\tau$ is close to 1, the assumption that only one process happens in one step does not hold, then the fitting result is no longer reliable. In contrast, the restriction on $\beta\tau$ is looser for our experiments, because the two-atom collision loss process cannot occur twice when the number of atoms is less than four. As shown in Fig.~\ref{fig2}, the fitting method works well when $\beta\cdot\SI{20}{\ms}=0.70$ which is close to 1.

\begin{figure}[!h]
\centering
\includegraphics[width=0.9\linewidth]{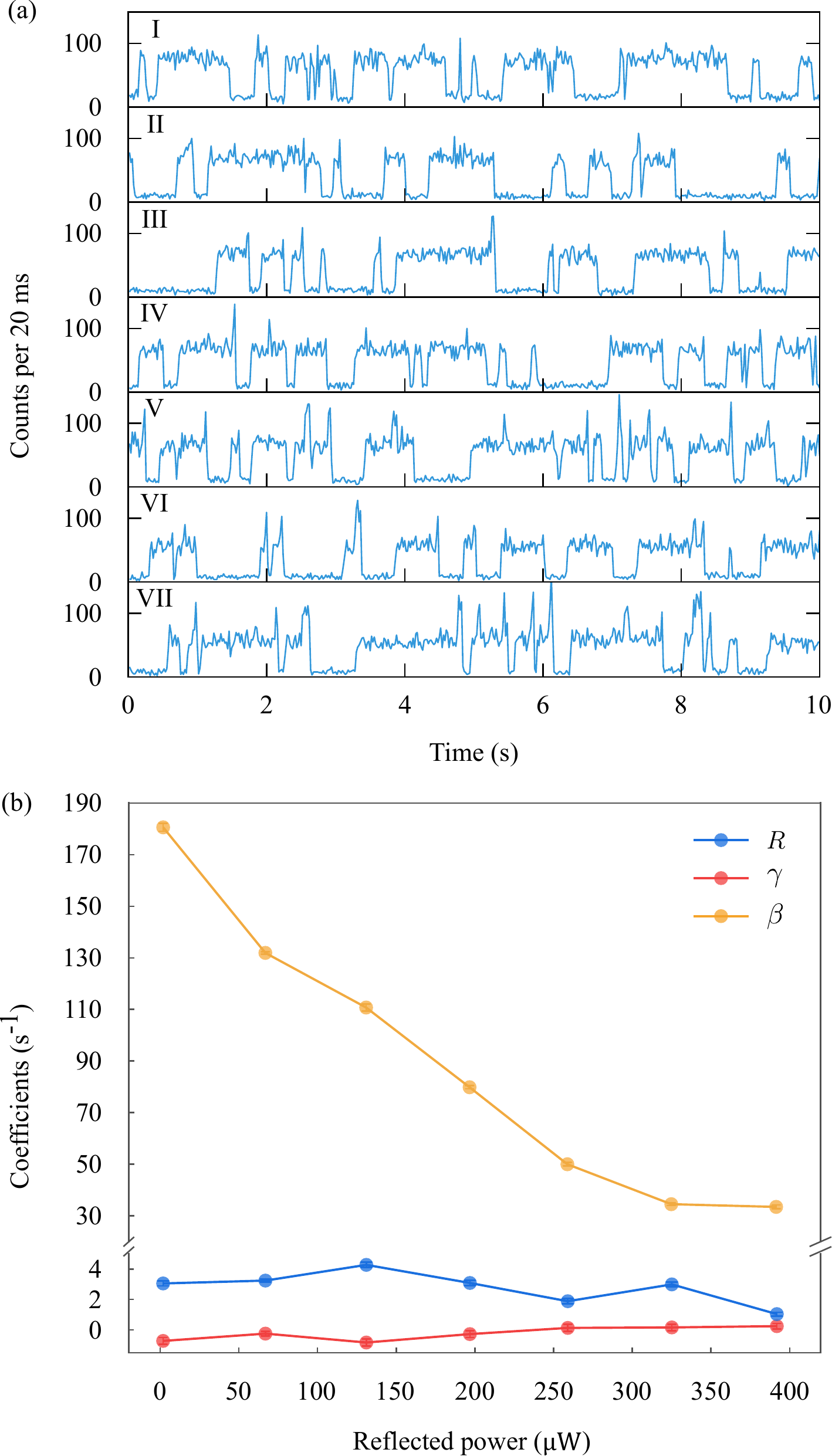}
\caption{(a) Typical traces of the photon counting rates under different reflected powers of the trap beam, which are (\uppercase\expandafter{\romannumeral1}) \SI{2}{\uW}, (\uppercase\expandafter{\romannumeral2}) \SI{67}{\uW}, (\uppercase\expandafter{\romannumeral3}) \SI{131}{\uW}, (\uppercase\expandafter{\romannumeral4}) \SI{197}{\uW}, (\uppercase\expandafter{\romannumeral5}) \SI{259}{\uW}, (\uppercase\expandafter{\romannumeral6}) \SI{325}{\uW}, (\uppercase\expandafter{\romannumeral7}) \SI{392}{\uW}. (b) The fitted coefficients $R$, $\gamma$ and $\beta$ of the experimental configurations with different reflected powers. The error bars are the fitting errors corresponding to two standard deviations, which are smaller than the data point size.}
\label{fig3}
\end{figure}

Comparing the experimental data [Figs.~\ref{fig2}(a) and (c)] with the simulation results [Figs.~\ref{fig2}(b) and (d)], the loading rate in Fig.~\ref{fig2}(a) is slower than that in Fig.~\ref{fig2}(c). This is because the loading rate $R$ fluctuates in practice due to the inhomogeneous and fluttering atom cloud. It is also worth mentioning that both histogram peaks of zero and single atoms from the experiments in Fig.~\ref{fig2}(c) are shorter and wider than the fitted Poisson distributions, which also results from the fluctuation of the intensity of the cooling laser. The histogram of the simulation results, however, fits the Poisson distribution better [Fig.~\ref{fig2}(d)]. Another feature is that in Figs.~\ref{fig2}(c), the distribution between peaks is much higher than expectation, even for the simulation result [Figs.~\ref{fig2}(d)]. This is because the photon counts is accumulated in \SI{20}{\ms}, thus once the atomic number changes in this \SI{20}{\ms}, the photon counts will be a mixture of two states. The probability of the jump is equal for any moment in \SI{20}{\ms}, so the photon counting rate distribution between peaks will present as a uniform background. The height of this uniform background mostly depends on the loading rate. When $R$ becomes larger, the jump happens more frequently and hence the background increases.

Now with the validated model, we can systematically investigate the atomic collision dynamics in the modified dipole trap. A tunable optical attenuator is placed between Lens~2 and the CCD to change the power of the backward dipole laser reflected from the CCD, which determines the modulation depth of the trap potential. When the reflected power increases, the two-atom state occurs more frequently, as shown in Fig.~\ref{fig3}(a). The fitted $\beta$ varies from $180.61$ to $33.46~\mathrm{s}^{-1}$ with the increase in the reflected power from $2$ to \SI{392}{\uW} [Fig.~\ref{fig3}(b)]. In contrast, $R$ and $\gamma$ remain nearly unchanged, because the only varying experimental parameter is the reflected power. Other configurations, such as the effective size of the dipole trap, the temperature and density of the atom cloud, and the vacuum pressure, are fixed. Note that a negative $\gamma$ [as the blue line in Fig.~\ref{fig3}(b)] is not attainable in physics, which is attributed to a small enough single-atom loss rate $\gamma$ with a restricted precision. Moreover, the typical trace with a small $\beta$ [tace \uppercase\expandafter{\romannumeral7} of Fig.~\ref{fig3}(a)] shows that the two-atom event occurs when there is already one atom in the trap, indicating the one-by-one continuous loading of atoms in the experiments. We also noticed that a two-atom event usually ends with a zero-atom state, which can be explained by the red-detuned light-assisted two-atom collision~\cite{Schlosser2002,Fung2015,Grunzweig2010a}. The jump down from the two- to zero-atom state also manifests that the dominant atom-loss mechanism for the two-atom state is two-atom collision instead of single-atom loss ($\beta\gg2\gamma$), which is in accord with our fitting parameters shown in Fig.~\ref{fig3}(b). In trace \uppercase\expandafter{\romannumeral7} of Fig.~\ref{fig3}(a), the longest lifetime and the average lifetime of the two-atom state detected are \SI{180}{\ms} and $\SI{28.2}{\ms}$ respectively, where, with the parameters fitted from the experiment, the predicted average lifetime from our model is $\SI{29.4}{\ms}$.

As demonstrated above, the modulation of the dipole trap allows the continuous and independent control of the two-atom collision rate. It is known that the single-atom loading rate $R$ and the single-atom loss rate $\gamma$ can also be adjusted independently. For example, $R$ is determined by the density of the atom cloud, which can be tuned by varying the current in the anti-Helmholtz coils~\cite{Schlosser2001}. Change of $\gamma$ is also possible to be obtained via the cooling laser parameters by affecting the polarization gradient cooling process~\cite{Chin2017,Shang1990}. Combining the control of these parameters, it is possible to load a given number of atoms in a single dipole trap. As predicted by our model, a probability of a three-atom state can be as high as $26\%$ when $R=50\,s^{-1}$ and $\beta=6\,s^{-1}$ with a corresponding lifetime of $\sim14.7\,\mathrm{ms}$, and a probability of a four-atom state of $23\%$ when $\beta$ changes to $4\,s^{-1}$ with a lifetime of $\sim13.5\,\mathrm{ms}$.

In conclusion, we have experimentally controlled atom collision and demonstrated two-atom trapping in a tight optical dipole trap by introducing a backward propagating dipole laser. A theoretical model is developed to numerically extract the loading, loss and collision rate, which provides a quantitative approach to study the dynamics of atoms in a dipole trap. The collision rate is tuned from $180.61~\mathrm{s}^{-1}$ to $33.46~\mathrm{s}^{-1}$, resulting in a two-atom trapping lifetime of $29.9\,\mathrm{ms}$. Our work provides an effective approach for realizing multiple atom trapping and manipulation in a single dipole trap setup, which promises the capture of a few atoms for collective light-matter interactions~\cite{Glicenstein2020} and few-atom ensemble coding~\cite{Pedersen2009a} and paving the way of the study of quantum optics and quantum information.

\noindent \textbf{\large{}Acknowledgment}{\large\par}

\noindent This work was funded by the National Key R \& D Program (Grant No.~2021YFA1402004) and the National Natural Science Foundation of China (Grants No.~11922411, U21A20433, and U21A6006). CLZ was also supported by the Fundamental Research Funds for the Central Universities.
%The numerical calculations in this paper have been done on the supercomputing system in the Supercomputing Center of University of Science and Technology of China.
This work was partially carried out at the USTC Center for Micro and Nanoscale Research and Fabrication.

% Bibliography

%\bibliographystyle{Zou}

%\bibliography{collision}

%Control: production of eprint (-1) disabled

%Control: production of eprint (-1) disabled
%

\end{document}